\documentclass[12pt]{iopart}

\usepackage{color}
\usepackage{latexsym}
\usepackage{graphics}
\usepackage{subfigure}
\usepackage{epsfig}
\usepackage{bbm}
\begin{document}

\newcommand{\aop}{\hat{a}}
\newcommand{\adop}{\hat{a}^{\dag}}
\newcommand{\nop}{\hat{n}}

\newcommand{\beq}{\begin{equation}}
\newcommand{\eeq}{\end{equation}}

\newcommand{\bea}{\begin{eqnarray}}
\newcommand{\eea}{\end{eqnarray}}

\newcommand{\Bra}[1]{\ensuremath{\langle#1|}}
\newcommand{\bra}[1]{\ensuremath{\langle#1|}}
\newcommand{\Ket}[1]{\ensuremath{|#1\rangle}}
\newcommand{\ket}[1]{\ensuremath{|#1\rangle}}
\newcommand{\BraKet}[2]{\ensuremath{\langle #1|#2\rangle}}
\newcommand{\braket}[2]{\ensuremath{\langle #1|#2\rangle}}
\newcommand{\ew}[1]{\ensuremath{\langle #1\rangle}}

\newcommand{\sigmahat}{\ensuremath{\hat{\sigma}}}
\newcommand{\jhat}{\ensuremath{\hat{\jmath}}}
\newcommand{\subj}{\ensuremath{\hat{\bar{\imath}}}}

\newcommand{\LN}{{\rm LN}}
\newcommand{\todo}[1]{{\color{red} #1}}

\newcommand{\ie}{{\it i.e.}}
\newcommand{\eg}{{\it e.g.}}
\newcommand{\hil}{\mathcal{H}}

\newcommand{\kB}{k_\textrm{B}}

\newcommand{\z}{\zeta}

\newcommand{\HBBH}{H_\textrm{BBH}}
\newcommand{\HSBm}{H_\textrm{eff}}

\newcommand{\Ub}{U_\textrm{b}}
\newcommand{\UB}{U_\textrm{B}}
\newcommand{\UbB}{U_\textrm{bB}}
\newcommand{\mub}{\mu_\textrm{b}}
\newcommand{\muB}{\mu_\textrm{B}}
\newcommand{\Jb}{J_\textrm{b}}
\newcommand{\JB}{J_\textrm{B}}

\renewcommand{\b}{\mathbf{b}}
\newcommand{\B}{\mathbf{B}}

\newcommand{\n}{\mathbf{n}}
\newcommand{\N}{\mathbf{N}}

\newcommand{\compB}{\mathbf{\mathcal{B}}}
\newcommand{\compN}{\mathbf{\mathcal{N}}}
\newcommand{\compV}{\mathcal{V}}
\newcommand{\eins}{\ensuremath{\mathbbm 1}}

\newcommand{\Sx}{\mathbf{S}^x}
\newcommand{\Sy}{\mathbf{S}^y}
\newcommand{\Sz}{\mathbf{S}^z}
\newcommand{\Salpha}{\mathbf{S}^\alpha}
\newcommand{\Sbeta}{\mathbf{S}^\beta}
\newcommand{\Sgamma}{\mathbf{S}^\gamma}
\newcommand{\field}{\mathbb C}
\newcommand{\veq}{\simeq}
\newcommand{\calhanc}{ {\cal H}_A \otimes {\cal H}_{Aa} \otimes {\cal
    H}_B \otimes {\cal H}_{Ba}}
   
\newcommand{\via}{{\it via}\ }

\title[Transport and Entanglement Generation in the Bose-Hubbard Model]{Transport and Entanglement Generation in the Bose-Hubbard Model}

\author{O. Romero-Isart$^1$, K. Eckert$^1$, C. Rod\'o$^1$, and A. Sanpera$^{1,2}$}

\address{$^1$ Grup de F\'isica Te\`orica, Departament de F\'isica, Universitat Aut\`onoma de Barcelona, E-08193 Bellaterra, Spain.}
\address{$^2$ Instituci\'o Catalana de Recerca i Estudis Avan\c cats (ICREA).}

\eads{\mailto{ori@ifae.es}, \mailto{kai@ifae.es},
\mailto{rodo@ifae.es}, and \mailto{sanpera@ifae.es}}

\begin{abstract}
We study entanglement generation via particle transport across a one-dimensional
system described by the Bose-Hubbard Hamiltonian. We analyze how the competition between interactions and tunneling 
affects transport properties and the creation of entanglement in the occupation number basis. Alternatively, we propose
to use spatially delocalized quantum bits, where a quantum bit is defined by
the presence of a particle either in a site or in the adjacent one.
Our results can serve as a guidance for future experiments to characterize entanglement of ultracold gases in one-dimensional optical lattices.
\end{abstract}

\section{Introduction}

The generation of entanglement between distant nodes of a quantum network  has profound implications for quantum computation and information \cite{bennett:2000}, and has triggered a remarkable effort into the study of entanglement generation and quantum state transport between distant lattice sites (see, e.g., Refs. \cite{bose:2003,christandl:2004,romeroisart:2006,cubitt:2007}  for discussions of transport in spin chains; entanglement dynamics in such systems is, e.g., studied in \cite{montangero:2003,jvidal:2004,amico:2004}). Furthermore, the recent experimental achievements in loading either bosonic and/or fermionic ultracold atomic gases into optical lattices, which permits to reproduce very accurately several spin Hamiltonians, have spurred an enormous interest in lattice systems (see \cite{lewenstein:2006} and references therein). 
 
Here we focus on one of the simplest but yet non-trivial lattice model, the so-called one-dimensional (1D) Bose-Hubbard Model (BHM) \cite{haldane:1980,fisher:1989}. It describes a system of spinless  bosons with (repulsive) on-site interaction, 
which can hop (tunnel) between adjacent sites of a 1D lattice. In general, this genuinely many-body model cannot be reduced to an effective non-interacting one. As proposed in \cite{jaksch:1998} and later experimentally demonstrated \cite{greiner:2002}, it can be realized by confining ultracold atoms in an optical lattice. In this article we study transport properties in this system. Here, by transport
we refer to the dynamics obtained when an extra particle is loaded onto a system that previously was cooled to its ground state. 
In some limiting cases, transport can be described in the language of continuous time quantum walks (a quantum analogue of
classical random walks) and closed analytical expressions can be found \cite{bose:2006}. When competition between
hoping and on-site interactions arises, the transport properties as well as 
the generation of entanglement between distant locations of the lattice is substantially modified.

The article is structured as follows: in the present section we review the essential 
properties of the Bose-Hubbard model. As one limiting case we 
identify transport across the lattice with continuous time quantum walks.
We discuss quantification and characterization of entanglement 
for a single particle propagating in a 1D lattice. 
In Sec.~\ref{sec:trans}, we analyze the generation of entanglement 
between distant lattice sites when an extra particle is loaded on top of the ground state. In
Sec.~\ref{sec:sdq}, we discuss entanglement in the so-called Spatially Delocalized
Qubit (SDQ) basis and investigate the role of interactions within this scheme.
Finally, we summarize briefly in Sec.~\ref{sec:concl}.

\subsection{The Bose-Hubbard Hamiltonian}

The Bose-Hubbard Hamiltonian for a 1D lattice of $M$ sites
(with open-boundary conditions) has the form
\begin{equation}\label{Hamiltonian BH}
\hat{H}_{\rm BH}=-J\sum_{i=1}^{M-1} \left(\adop_i \aop_{i+1} + \adop_{i+1} \aop_{i}
 \right) +\sum_{i=1}^M \epsilon_i \nop_i+
\frac{U}{2}\sum_{i=1}^M \nop_i (\nop_i-1),
\end{equation}
where $\aop_i$ and $\adop_i$ are the bosonic annihilation
and creation operators for a particle on the $i$-th lattice site, $\nop_i=\adop_i \aop_i$ is the corresponding bosonic number operator, 
and $\epsilon_i$ accounts for the single-particle on-site energy. 
The Bose-Hubbard model assumes only nearest-neighbour tunneling with constant amplitude $J$ and pairwise
interaction between bosons on the same site leading to an energy shift $U$.

A particularly clean realization of such a Hamiltonian is obtained by trapping
neutral bosons {\it via} the dipole force in an optical lattice \cite{jaksch:1998}. By taking
the trapping sufficiently tight in two directions (say $y$ and $z$),
an effective one-dimensional system can be realized. The corresponding
Hamiltonian in second quantization notation reads
\begin{eqnarray}\label{eq:hamfree}
\hat\hil_{\rm OL}&=&\int 
dx\,\hat{\psi}^{\dagger}(x)\left(\frac{p_x^2}{2m}+V_0\sin^2(\pi x/d)\right)\hat{\psi}(x)+
g\int dx\,\hat{\psi}^{\dagger}(x)\hat{\psi}^{\dagger}(x)
\hat{\psi}(x)\hat{\psi}(x)\nonumber\\
&\equiv&\int dx\,\hat{\psi}^{\dagger}(x)H_{\rm free}\hat{\psi}(x)+\hat{\hil}_{\rm int}.
\end{eqnarray}
The optical lattice is completely characterized by its depth $V_0$, that can be controlled through the laser intensity, and
by the wave number ${\it k}= 2 \pi/ \lambda$ where $d=\lambda/2$ is the lattice periodicity. The effective 1D 
interaction strength $g=2\pi\hbar a_s\omega_t$ (where $\omega_t$ is the transversal frequency of the trap assumed equal in the $y,z$ directions and
$a_s$ is the atomic scattering length) can be changed either by modifying the confinement of the atoms in the two orthogonal directions or, alternatively, \via a Feshbach resonance, which even allows to change the sign of $a_s$ \cite{cornish:2000}.
For periodic boundary conditions (or large enough lattices) 
the bosonic operators can be expanded in terms of Bloch functions.
In the low temperature regime and with typical bosonic interaction strengths, excitations to higher bands  can be neglected for sufficiently deep lattices. 
The dynamics is then restricted to the lowest Bloch band and the field operators can be expanded in terms of single-particle Wannier functions localized at each lattice site $x_i=d\,i$ as  $w(x-x_i)\equiv\braket{x}{w_i}$. The bosonic creation operators $\adop_i$ are now defined {\it via} $\ket{w_i}=\adop_i\ket{\Omega}$, $\ket{\Omega}$ being the vacuum. Eq.~(\ref{Hamiltonian BH}) is derived from Eq.~(\ref{eq:hamfree}) by
keeping only nearest-neighbor hopping and restricting the interactions between bosons to a contact (zero-range) potential. 
Under these approximations, the tunneling amplitude between adjacent sites reads $J=\bra{w_i}H_{\rm free}\ket{w_{i+1}}$,
and the on-site boson-boson interaction is given by $U=g\int dx |w(x)|^4$.

The Bose-Hubbard ground state with filling factor $\bar n$ is obtained by minimizing $\ew{\hat{H}_{\rm BH}-\mu\sum_i \hat n_i}$,
where the chemical potential $\mu$ fixes the total number of particles.
In the limit $U/J\rightarrow0$ (strictly speaking, in the thermodynamic limit for
$U/J<(U/J)_c \approx 3.44$ for $\bar n=1$ \cite{kuehner:2000}) it is energetically favorable 
to spread each particle over the whole lattice. For periodic boundary conditions,
a ground state with filling factor $\bar n$ can be explicitly written as
\begin{equation}
\ket{\psi^{\bar n}_{\rm GS,SF}}= \frac{1}{ \sqrt{(M\bar n)!}}\left(\frac{1}{\sqrt{M}}\sum_i^M\adop_i\right)^{{M\bar n}}\ket{\Omega},
\end{equation}
where $M\bar n$ is the total number of particles.
This {\rm superfluid} (SF) state is characterized by large fluctuations of the on-site
number of particles, divergent correlation length, and a vanishing gap.
In the opposite limit, for $U/J\rightarrow\infty$, the ground state is a
{\rm Mott Insulator} (MI) state, \ie, a product state with
well-defined number $\bar n$ of atoms per site
\begin{equation}
\ket{\psi^{\bar n}_{\rm GS,MI}}=\frac{1}{\sqrt{{\bar n}!}}\prod_i^M
\left(\adop_i\right)^{\bar n}\ket{\Omega},
\end{equation}
where $\bar n-1<\mu/U<\bar n$. The MI state has finite correlation length and 
a gapped spectrum. Increasing $U/J$ from $0$ to $\infty$ for {\it integer} filling factor (at $T=0$ and in the infinite system), 
the 1D-Bose-Hubbard model undergoes a quantum phase transition (which corresponds to a Kosterliz-Thouless phase transition).
There is also a generic phase transition which is crossed when the value of $U/J$ is fixed and   
the chemical potential $\mu$ changes. In this case, the number of particles is not conserved and the behavior of the ground state
near the phase transition simply corresponds to a weakly interacting condensate with a non-integer filling factor.
See \cite{lewenstein:2006} and references therein for a detailed review of the properties of the Bose-Hubbard model.

\subsection{Continuous time quantum walks}
 
For the simplest transport case in the Bose-Hubbard model, namely, for a single boson placed in an otherwise empty lattice,
the dynamics is equivalent to the one of a free particle moving in (finite) discretized one-dimensional space. 
Recently, such a model and its generalizations to more complex underlying graphs have been studied intensively in the context of 
Continuous Time Quantum Walks (CTQWs). Quantum walks, either continuous or discrete, have been proposed
as quantum versions of classical random walks and analyzed with the aim of 
constructing new types of algorithms. They have also been studied, e.g., in relation to decoherence properties of lattice systems. 
See \cite{kempe:2003} (and references therein) for an excellent review of the topic. 

\begin{figure}[t]
\begin{center}
\centering
\includegraphics[width=0.7\linewidth]{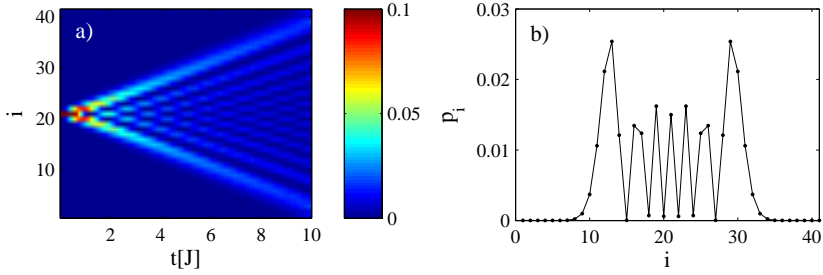}
\end{center}
\caption{
(a) Probability distribution $|c_i(t)|^2$ versus time for a CTQW starting at site $(M+1)/2$ in a chain of $M=41$ sites; 
(b) probability distribution at time $t=5 J$.}
\label{fig:ctqw}
\end{figure}

The definition of a CTQW is closely related to
{\it classical continuous time random walks} \cite{childs:2002,kempe:2003}. Let us consider 
the classical situation of a ``particle" which can move on a set of vertices. 
The probability to jump from vertex $i$ to another vertex $j$ per unit time is denoted as $J_{ij}$, with $J_{ij}>0$ if both vertices are connected and
$J_{ij}=0$ otherwise. To conserve the total probability we demand $J_{ii}=-\sum_{j\neq i}J_{ij}$. If $p_i(t)$ is the probability
of being at time $t$ at vertex $i$, then
\beq
\frac{d p_i(t)}{dt}=\sum_j J_{ij}p_j(t).\label{eq:ctrw}
\eeq
Given a quantum state $\ket{\psi}$ in the Hilbert space $\hil=\left\{\ket{i},\,i=1\ldots M\right\}$
spanned by the vertices, the similarity between Eq.~(\ref{eq:ctrw}) and the Schr{\"o}dinger
equation for the amplitudes $\braket{i}{\psi(t)}$,
\beq\label{eq:hctqw}
{\rm i} \frac{d\:\braket{i}{\psi(t)}}{dt}=\sum_j\bra{i}\hat{H}\ket{j}\braket{j}{\psi(t)},
\eeq
suggests to define a quantum analogue of the classical random walk by identifying $J_{ij}$ with
the Hamiltonian matrix elements: $H_{ij}=\bra{i}\hat{H}\ket{j}=J_{ij}$. For 
vertices arranged on a finite line with $M$ sites and constant nearest-neighbour transition
probabilities $J_{ij}\equiv J$ in the graph,
equation (\ref{eq:hctqw}) is equivalent to the Schr{\"o}dinger equation of a
single particle evolving under the Bose-Hubbard Hamiltonian (\ref{Hamiltonian BH}).
To generate entanglement in an effective way, transport should be symmetric, thus
we impose an odd number $M$ of sites. Placing the particle in the middle of the chain, \ie, preparing
$\ket{\psi(t=0)}=\adop_{(M+1)/2}\ket{\Omega}$, dynamics after a time $t$ leads to the state
\beq
\ket{\psi(t)}=\sum_{i=1}^M c_i(t)\adop_i\ket{\Omega},
\eeq
where the coefficients $c_i(t)$ are given by \cite{bose:2003,christandl:2004}
\beq
c_i(t)=\frac{2}{M+1}\sum_{k=1}^{M}\left[\sin\left(\frac{\pi k}2\right)\sin\left(\frac{\pi k i}{M+1}\right)\right]
\exp\left[2{\rm i}Jt\cos\left(\frac{k\pi}{M+1}\right)\right].\label{eqn:coft}
\eeq
The probability distribution $p_i(t)=|c_i(t)|^2$, see Fig.~\ref{fig:ctqw}, is symmetric with
respect to the point $(M+1)/2$. For times $t$ such that $p_1(t)=p_M(t)\ll1$,  its standard deviation
$\Delta =\sqrt{\sum_i p_i i^2-(\sum_ip_i i)^2}$ grows linearly in time:
$\Delta \propto t$ \cite{kempe:2003}. This is in strong contrast to a classical 1D random walk, where $\Delta\propto\sqrt{t}$.

\subsection{Characterizing entanglement}\label{sec:entan}

Here we first discuss shortly how to characterize entanglement distributed in a CTQW. 
We will later generalize this discussion to the transport of a 
defect,  namely an extra particle, loaded on top of the
ground state of the Bose-Hubbard Hamiltonian. Since particles in the CTQW 
as well as in the spinless Bose-Hubbard model have no {\it internal} degrees of freedom, 
the notion of quantum bit (or quantum dit) and its extensions to entangled states 
and entanglement measures have to be redefined.
S.~Bose has discussed the distribution
of entanglement in the CTQW for a single particle as well as for $N$ non-interacting
indistinguishable bosons \cite{bose:2006} (corresponding to $N$ [symmetrized] versions of the quantum walk),
quantifying entanglement between the two outer lattice sites (say $1,\,M$) in the occupation number basis. 
To this aim, the reduced density matrix of sites $1$ and $M$ is expanded in the basis
$\{\ket{n_1,n_M}\}$, where $n_i$ corresponds to the number of particles at site $i$. Within this
basis, usual entanglement measures can be applied. 
Quantifying entanglement in this ``second quantized" formalism 
was introduced by P.~Zanardi \cite{zanardi:2002} and subsequently intensively discussed in the literature \cite{gittings:2002, omar:2002,simon:2002,shi:2003}. 
For just a single particle, mapping its presence (absence) on site $\it{i}$ to 
spin-up (spin-down) reduces the propagation of a single boson in the lattice to the dynamics of a single, initially localized, 
excitation in the $XY$ spin-chain \cite{bose:2003,christandl:2004}.
This analogy between ``space" and ``spin-$1/2$" entanglement is, however, limited:
a spin at site $i$ can be in any superposition  of ``up'' and ``down''; $\ket{\rm up}_i+\ket{\rm down}_i$, 
while $\ket{n_i=0}+\ket{n_i=1}$ has no physical meaning for massive particles. Also, in the occupation number basis, operators
must correspond to a direct sum of operators acting in sectors with fixed number
of particles. For these reasons, it is not clear whether entanglement in the
occupation number basis can be identified 
at all with usual ``spin"-entanglement.  In 
particular, there are no protocols that convert a (highly) entangled state in the occupation number basis to
a ``spin"-entangled state containing the same amount of entanglement%
\footnote{
In \cite{bose:2006} a method to convert  ``space" to ``spin" entanglement is given
for the coupled chain. It requires interaction between spins and final
measurements over intermediate sites. Thus it is not a local conversion scheme. Furthermore, it
allows to extract only up to one ebit. In \cite{gittings:2002}, a local protocol is presented to
extract an ebit from $(\ket{1,0}+\ket{0,1})/\sqrt2$, which however needs a sink/source for
particles.}.

Despite its practical drawbacks,
we investigate first the generation of entanglement in the occupation number basis, thereby merely using it
as a tool to characterize transport in the system.
For a single particle (\ie, the CTQW), the reduced density matrix of sites $1$ and $M$ is always
of the form
\beq
\hat{\rho}_{1M}(t)=\hat{\rho}^{(0)}\oplus\hat{\rho}^{(1)},\quad\hat{\rho}^{(0)}=1-2p_1(t),\quad\hat{\rho}^{(1)}=2p_1(t)\ket{\Psi^+}\bra{\Psi^+},
\label{eq:ln_p0}
\eeq
where the upper indexes correspond to the total particle number, and
$\ket{\Psi^+}=(\ket{1,0}+\ket{0,1})/\sqrt2$. To later be able to generalize to states with more 
than one particle, we measure the entanglement through the logarithmic negativity $\LN$ \cite{plenio:2007}. From Eq.~(\ref{eq:ln_p0}),
\bea
&\,&\LN(\hat{\rho}_{1M}(t))=\log_2 ||\hat{\rho}^{\Gamma}_{1M}(t)||_1\nonumber \\
&\,&=\log_2\left[ 
2p_1(t) + \sqrt{ \frac {\alpha+(1-2p_1(t))\sqrt{\beta}}{2}}
+\sqrt{\frac{\alpha-(1-2p_1(t))\sqrt{\beta}}{2}}\right]\label{eq:ln_p1},
\eea
where $\alpha=1-4p_1(t)+6p_1(t)^2,\,\beta=1-4p_1(t)+8p_1(t)^2$. 
Since we are considering open boundary conditions, the logarithmic negativity presents several
local maxima in time (as well as periodic revivals) due to multiple reflections. In order to
characterize the CTQW and subsequently the Bose-Hubbard model through the generated entanglement,
we consider only its first maximum. 


\section{Transport properties of the Bose-Hubbard model: from the Mott Insulator to the superfluid phase}\label{sec:trans}

\subsection{Transport of an additional particle on a Bose-Hubbard ground state background}

We now generalize the previous discussion transport of a single
particle in an empty lattice system to an extra particle propagating on a
background given by a Bose-Hubbard ground state.
Let us start by discussing the two extreme cases 
(i) $U/J\rightarrow\infty$ (Mott insulator phase) and (ii) $U/J\rightarrow0$ (superfluid phase).

\indent{\it (i) Mott Insulator phase}. 
Adding an extra-particle at site $(M+1)/2$ to the MI ground state 
with an integer filling factor $\bar n$ leads to the initial state  
\beq
\ket{\psi(t=0)}=\frac{\adop_{(M+1)/2}}{\sqrt{\bar n+1}}\ket{\psi_{\rm GS,MI}^{\bar n}}=
\frac{\adop_{(M+1)/2}}{\sqrt{\bar n+1}}\left(\frac{1}{\sqrt{{\bar n}!}}\right)^M\prod_{i=1}^{M}(\adop_i)^{\bar n}\ket{\Omega}
\eeq
For $U/J$ large, during time evolution the system will remain in the subspace spanned by states
$\{\ket{i_{\bar n}}=\adop_{i}\ket{\psi_{\rm GS,MI}^{\bar n}}/\sqrt{\bar n+1},i=1\ldots M\}$.
Noticing that 
$(-J\adop_{i+1}\aop_{i})\adop_{i}\ket{\psi_{\rm GS,MI}^{\bar n}}=-J({\bar n}+1)\adop_{i+1}\ket{\psi_{\rm GS,MI}^{\bar n}}$,
we find that the effective Hamiltonian, up to corrections
of order $(U/J)^{-1}$, reads
\beq
H_{\rm eff}=-J({\bar n}+1)\sum_i\left[\ket{i_{\bar n}}\bra{(i+1)_{\bar n}}+\ket{(i+1)_{\bar n}}\bra{i_{\bar n}}\right].
\eeq
Thus, $\ket{\psi(t)}=\sum_ic_i^{\bar n}(t)\ket{i_{\bar n}}$, with $c_i^{\bar n}=c_i(({\bar n}+1)t)$, \ie,
from bosonic enhancement the propagation is ${\bar n}+1$ times faster than in the pure CTQW, 
while the magnitude of the distributed entanglement is as before.\\

\indent{\it (ii) Superfluid phase}. In the opposite limiting case, for $U/J=0$,
approximating the ground state of the system with open boundaries 
by the one for periodic boundary conditions, the initial state reads
\beq
\ket{\psi(0)}=\alpha\adop_{(M+1)/2}\ket{\psi^{\bar n}_{\rm GS,SF}}=\alpha \frac{\adop_{(M+1)/2}}{\sqrt{(M\bar n)!}} \left[\frac{1}{\sqrt{M}}\sum_{i=1}^M\adop_i\right]^{M\bar n}\ket{\Omega},
\eeq
where $\alpha$ is a normalization constant. Evolution of this state leads to
\beq
\ket{\psi(t)}=\frac{\alpha}{\sqrt{(M \bar n)!}}\sum_{i=1}^M c_i(t)\adop_i\ket{\psi^{\bar n}_{\rm GS,SF}},
\eeq
since $-J[\sum_{i}(\adop_{i+1}\aop_{i}+\adop_{i-1}\aop_{i}),\adop_j]=-J(\adop_{j+1}+\adop_{j-1})$.
Thus, the extra particle added on top of the superfluid ground state propagates as a 
single particle in an otherwise empty lattice.

\begin{figure}[t]
\centering
\includegraphics[width=0.8\linewidth]{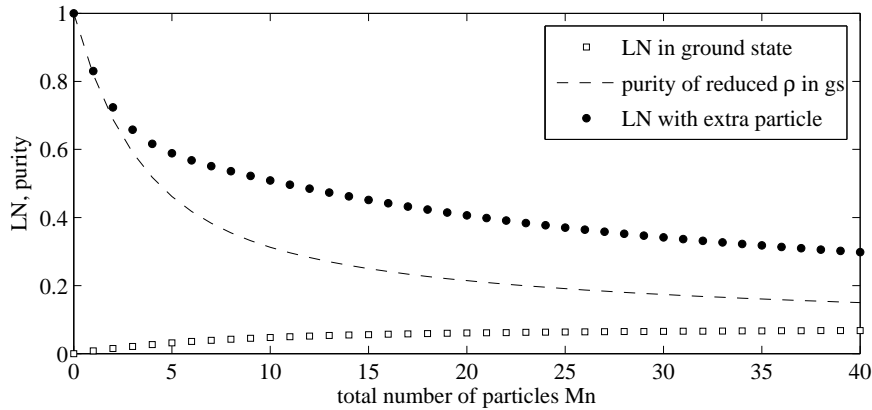}
\caption{
Entanglement (as measured by the logarithmic negativity $\LN$) between the outer sites for the superfluid ground
state (squares) and the superfluid ground state with an extra particle using $c_i=(\delta_{i1}+\delta_{iM})/\sqrt2$
(filled circles) as discussed in the text. The number of sites is $M=20$. The purity $\rm{tr}[(\hat \rho_{\rm SF})^2]$
of the reduced density matrix is plotted
as a dashed line. For the insulator case (with integer $\bar n$), always $\LN(\hat \rho'_{{\rm MI},\bar n})=0$ for the ground
state and $\LN(\hat \rho'_{{\rm MI},\bar n})=1$ for the ground state with an additional particle and $c_i$ as above.}
\label{fig:ent_gs_n}
\end{figure}

The entanglement between the two outer sites however is strikingly different for the Mott insulator and
the superfluid state, as we will discuss now. Let us first consider entanglement between the outer sites
just for the ground state, without any additional particle. In the insulator case, we have
\bea
\ket{\psi^{\bar n}_{\rm GS,MI}}&=&\frac{1}{(\sqrt{{\bar n}!})^M}\prod_{i=1}^{M}(\adop_i)^{\bar n}\ket{\Omega}
=\frac{1}{{\bar n}!}(\adop_1\adop_M)^{\bar n}\frac{1}{(\sqrt{{\bar n}!})^{M-2}}\prod_{i=2}^{M-1}(\adop_i)^{\bar n}\ket{\Omega}\\
&\equiv&\ket{\phi_{\rm MI}}_{1M}\otimes\ket{\psi_{\rm MI}}_{2\ldots M-1}.
\eea
The reduced density matrix of sites $1,\,M$ then reads
\beq
\hat \rho_{\rm MI,\bar n}={\rm tr}_{2\ldots M-1}(\ket{\psi^{\bar n}_{\rm GS,MI}}\bra{\psi^{\bar n}_{\rm GS,MI}})=
\ket{\phi_{\rm MI}}\bra{\phi_{\rm MI}}=\ket{\bar n,\bar n}\bra{\bar n,\bar n},
\eeq
where the latter expression is in the occupation number basis. $\hat \rho_{\rm MI\bar n}$ is a non-entangled pure state.
On the other hand, for the superfluid ground state and $M>2$,
\bea
\ket{\psi^{\bar n}_{\rm GS,SF}}&=&\frac{1}{\sqrt{(M{\bar n})!}}\left[\frac{1}{\sqrt{M}}\sum_{i=1}^M\adop_i\right]^{M\bar n}\ket{\Omega}\\\nonumber
&=&\sum_{k=0}^{M\bar n}\gamma_k
\frac1{\sqrt{k!}}\left(\frac{\adop_1+\adop_M}{\sqrt{2}}\right)^{k}
\frac1{\sqrt{(M\bar n-k)!}}\left(\frac1{\sqrt{M-2}}\sum_{j=2}^{M-1}\adop_j\right)^{M\bar n-k}\!\! \! \! \!  \ket{\Omega}\\
&\equiv& \sum_{k=0}^{M\bar n}\gamma_k\ket{\phi^{\,k}_{\rm SF}}_{1M}\otimes\ket{\psi^{\,k}_{\rm SF}}_{2\dots M-1}
\eea
with
\beq
\gamma_k=\sqrt{{{M\bar n}\choose k}\frac{2^{k}(M-2)^{M\bar n-k}}{M^{M\bar n}}}.
\eeq
The reduced density matrix reads
\beq
\hat \rho_{\rm SF,\bar n}=\sum_{k=0}^{M\bar n}\gamma_k^2\ket{\phi_{\rm SF}^{\,k}}\bra{\phi_{\rm SF}^{\,k}}.
\eeq
Except for the trivial cases $M\bar n=0$ or $M=2$, the state $\hat \rho_{\rm SF,\bar n}$ is mixed, has a direct
sum structure in the occupation number basis, and is entangled. As Fig.~\ref{fig:ent_gs_n} shows,
entanglement in the ground state grows (weakly) with the total number of particles $M\bar n$ in the system
(keeping the number of sites $M$ fixed). This happens despite the fact that
the purity of the reduced state decreases at the same time, meaning that the outer sites also become entangled
to the inner part of the chain. The reason is that adding more particles
corresponds to adding more degrees of freedom which can be entangled in the occupation number basis.
For this particular case, this leads to an increase in entanglement.

In order to discuss the entanglement generated from an extra particle on top of the ground state, let us consider the simplified situation $c_i=(\delta_{i1}+\delta_{iM})/\sqrt2$. For the Mott case,
\beq
\hat \rho'_{\rm MI,\bar n}=\frac14\left(\ket{\bar n+1,\bar n}+\ket{\bar n,\bar n+1}\right)\left(\bra{\bar n+1,\bar n}+\bra{\bar n,\bar n+1}\right),
\eeq
which is a pure state with logarithmic negativity $\LN(\hat \rho'_{{\rm MI},\bar n})=1$, independently of $\bar n$. Note that the entanglement is completely contained in the 
sector of $2\bar n+1$ particles shared between the outer sites. This is different for the superfluid phase. Here
the extra particle on top of the ground state with filling factor $\bar n$ leads to 
\beq
\hat \rho'_{\rm SF,\bar n}=\sum_{k=0}^{M\bar n+1}(\gamma'_k)^2\ket{\phi_{\rm SF}^{\,k}}\bra{\phi_{\rm SF}^{\,k}},
\eeq
with $\gamma'_0=0$, and
\beq
(\gamma'_k)^2=\frac{k(\gamma_{k-1})^2}{1+2\bar n}\quad\textrm{for}\quad k>0.
\eeq
Clearly, for an empty lattice, $\bar n=0$, the choice of $c_i$ leads to $\LN(\hat \rho'_{{\rm SF},\bar n=0})=1$. For $\bar n>0$, 
the logarithmic negativity $\hat \rho'_{\rm SF}$ decreases (Fig.~\ref{fig:ent_gs_n}).
Again, as $\bar n$ grows the number of degrees of freedom which potentially can be entangled increases also.
Still, in this case the two outer sites are less entangled due to the smaller purity of the reduced
density matrix (the outer sites are more entangled to the inner part of the chain).
It might seem counter-intuitive that
\beq
\LN(\hat \rho'_{\rm SF,\bar n})<\LN(\hat \rho'_{\rm SF,\bar n=0}),
\eeq
despite the additivity property of the logarithmic negativity \cite{plenio:2007}.
Here we should again remark, that the occupation number basis which is used to quantify entanglement does
{\it not} reflect the tensor product structure of individual particles. In fact, the bosonic particles themselves
live in a symmetrized subspace which does not have the structure of a full tensor product (this indeed is the
reason why for bosonic particles new measures of entanglement have to be introduced).

Let us finally note that if the systems dynamics is governed by the Bose-Hubbard Hamiltonian, then the
logarithmic negativity will always be below the values of Fig.~\ref{fig:ent_gs_n}, as in this case
the $c_i(t)$'s are given by Eq.~(\ref{eqn:coft}).

\begin{figure}[t]
\centering
\includegraphics[width=0.8\linewidth]{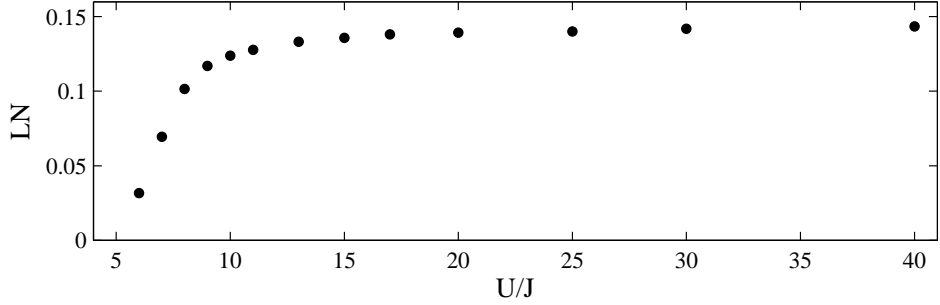}
\caption{
Entanglement between the two outer sites of the lattice, measured by the logarithmic negativity $\LN(\hat{\rho}_{1M})$, versus $U/J$ 
for a chain of $M=41$ sites (after inserting an extra particle at site at $(M+1)/2$).
The chemical potential $\mu$ is adapted to have a
ground state with mean number of particles per site $\bar n=1$ ($\mu=U/(2J)$ for $U\gg J$).
The data is obtained through MPS simulations truncating the basis to states of up to $5$
particles per site and using $D \le 20$.}
\label{fig:lnvsu}
\end{figure}
\begin{figure}[t]
\centering
\includegraphics[width=0.1\linewidth]{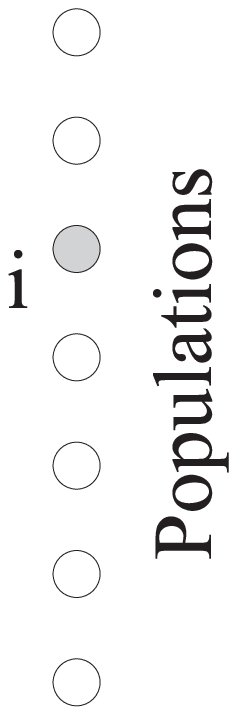}
\includegraphics[width=0.7\linewidth]{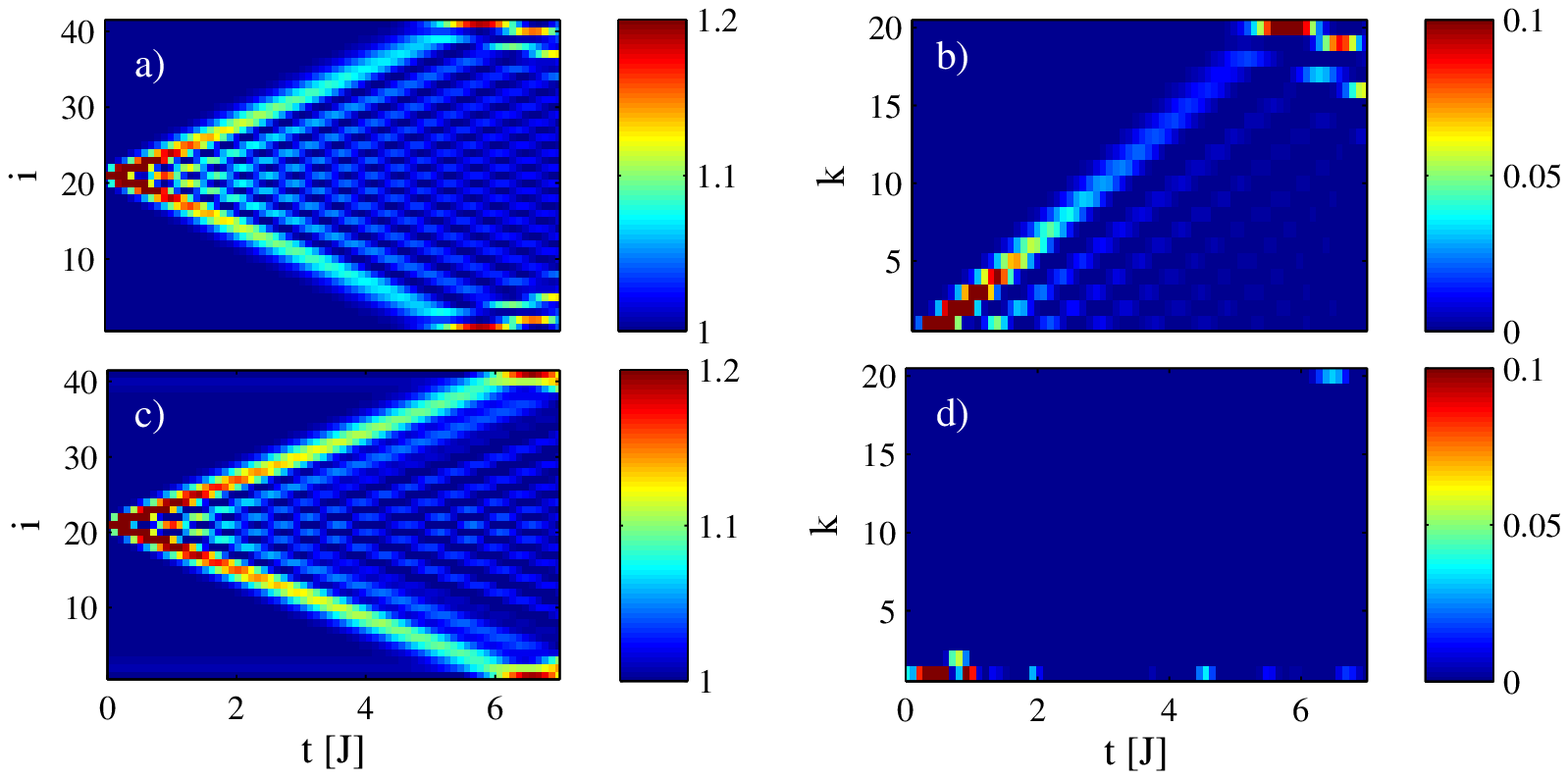}
\includegraphics[width=0.1\linewidth]{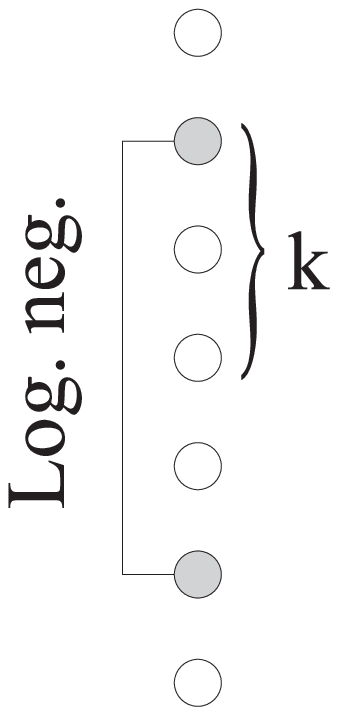}
\caption{
Left column: mean ocupation number $n_i=\langle\hat n_i\rangle$ of site $i$ as a function of time
for a chain of $M=41$ lattice sites ((a) $U/J=40$; (c) $U/J=6$). 
The system is initially cooled to its ground state with filling factor $\bar n=1$.
At $t=0$, a further particle is loaded at site $i=20$. Right column: logarithmic negativity $\LN(\hat{\rho}_{21-k,21+k})$
of the reduced density matrix of sites $21-k$ and $21+k$, $k=1,\ldots,20$ versus time for (b) $U/J=40$, (d) $U/J=6$.} 
\label{fig:sfmott}
\end{figure}

To study how the competition between interactions and tunneling affects the generation of entanglement
when adding an extra particle, we numerically calculate
time evolution of the initial state for a wide range of parameters,  
$6<U/J<40$. We use standard numerical MPS algorithms \cite{garcia-ripoll:2006}.
As we limit the number of particles per site to $5$ in our simulations, we cannot
study states well inside the superfluid regime, where for $\bar n=1$ up to $M$ particles
per site have to be taken into account.
Still, already for values of $U/J<15$ drastic changes due to the
possibility of tunneling are manifested in the entanglement between
the outer lattice sites. To calculate $\ket{\psi(t)}=\exp (-{\rm i}\hat{H}_{\rm BH} t)\ket{\psi^{\bar n=1}_{GS}}$, 
we first obtain the ground state of the Bose-Hubbard Hamiltonian for different
values of the parameter $U/J$. We choose the chemical potential as $\mu/J=U/(2J)$ in the case of large $U/J$. 
For small values of the ratio $U/J$, we adjust
$\mu$ appropriately, in order to obtain the ground state $\ket{{\Psi}^{\bar n=1}_{GS}}$ 
with filling factor $\bar n=1$\footnote{
In a one dimensional system the lobes of the Mott insulating phase are
much stronger deformed than in two or three dimensions \cite{freericks:1996}.
}.
Having the ground state, we add a particle to the system and obtain the dynamics through
a time-dependent MPS simulations. From the MPS state, the reduced density matrix $\hat \rho_{1,M}$ of the outer
sites can be extracted efficiently, such that we can compute the logarithmic negativity as a function of
time. In Fig.~\ref{fig:lnvsu} we display the value of $\LN (\hat{\rho}_{1,M})$ at
its first maximum as a function of $U/J$.
In Fig.~\ref{fig:sfmott} we analyze time evolution in detail for two cases: well in the MI phase ($U/J=40$) and
close to the SF phase ($U/J=6$). Figs.~\ref{fig:sfmott} (a,c) show the propagation of the excitation,
\ie, the mean occupation number $\ew{n_i}$ versus time. The propagation in the two cases is very similar
(as it is visible from the figures, the evolution is slower for $U/J=6$,
though not by a factor of $2$ as it would be the case
for $U/J=0$). The propagation of entanglement, visualized in Figs.~\ref{fig:sfmott} (b,d) through the logarithmic
negativities of the reduced density matrices of sites $(M-1)/2\pm k$, is different in the two cases.
Clearly, the efficient generation of entanglement in the occupation number basis requires
a Mott insulator background. As we demonstrated, entanglement generation is much less
efficient if tunneling becomes comparable to or larger than interactions.

A strategy to increase entanglement in a system of ideal (interactionless) bosons
in the occupation number basis consists in loading several bosons in a given site of an otherwise
empty lattice \cite{bose:2006}. For interacting particles, however, in the limit $U\gg J$,
entanglement does not increase if all $N$ particles are initially
located at the same site. In this case, tunneling of a single particle alone is strongly suppressed
and atoms tunnel together. Treating the tunneling term perturbatively, it can be seen that
the evolution is slower by a factor $J/(UN)$. 

\section{Creating entangled ``spatially delocalized quantum bits"}\label{sec:sdq}

In this section we discuss an alternative way of defining a quantum bit
in a lattice filled with spinless bosons. This definition does not rely on the occupation number basis,
leading to a notion of entanglement which is physically more sound.  
We use the concept of  ``spatially delocalized quantum bits'' (SDQs), in which the binary 
alternative consists in having an atom either in one side or in the adjacent one. 

To be specific, we use a single particle shared between two (adjacent) sites $i,\,i+1$ of
the chain to define a quantum bit by identifying $\ket{0}_{\rm SDQ}\equiv\ket{n_i=1,n_{i+1}=0}$ and
$\ket{1}_{\rm SDQ}\equiv\ket{n_i=0,n_{i+1}=1}$. Though in this definition
again a qubit is defined \via absence or presence of a particle on a lattice site,
now one particle is necessary {\it per} qubit. In this way, for any unitary transformation
the number of particles is conserved locally. For such ``charge" or ``spatially delocalized"
qubits, implementations of quantum gates have been proposed for bosonic atoms in
optical lattices \cite{mompart:2003}, as well as for electrons in
quantum dots \cite{renzoni:2001} or photons in photonic crystals \cite{angelakis:2004}.

To entangle two such qubits at the ends of a 1D chain of even length $M$, we place two 
particles at sites $M/2$ and $M/2-1$ (in the middle of the chain) and let the system evolve.
As before, we start by considering an otherwise empty chain, \ie, the quantum walk with two particles:
$\ket{\psi(t=0)}=\adop_{M/2}\adop_{M/2+1}\ket{\Omega}$.
For a spatially delocalized qubit defined through a particle shared between sites $i,j$, we introduce
the corresponding projection operators $\hat P^{(i,j)}_{\alpha}=\ket{\alpha}_{\rm SDQ}\bra{\alpha}$. For instance, 
\beq
\hat P^{(i,j)}_{0}=\ket{0}_{\rm sdq\!}\bra{0}=\ket{n_i=1,n_{j}=0}\bra{n_i=1,n_{j}=0}.
\eeq
$\hat P_0^{(i,j)}+\hat P_1^{(i,j)}$ thus projects onto the subspace having a single particle shared
between sites $i$ and $j$. If two qubits are defined on sites $(1,2)$ and $(M-1,M)$ respectively,
we obtain the density matrix for the two spatially delocalized qubits as
\beq
\hat{\rho}^{\rm sdq}_{\alpha\beta,\alpha',\beta'}(t)=
\hat P^{(1,2)}_{\alpha}\otimes \hat P^{(M-1,M)}_{\beta}\ket{\psi(t)}\bra{\psi(t)}\hat P^{(1,2)}_{\alpha'}\otimes \hat P^{(M-1,M)}_{\beta'}.
\eeq
\begin{figure}[t]
\centering
\includegraphics[width=0.8\linewidth]{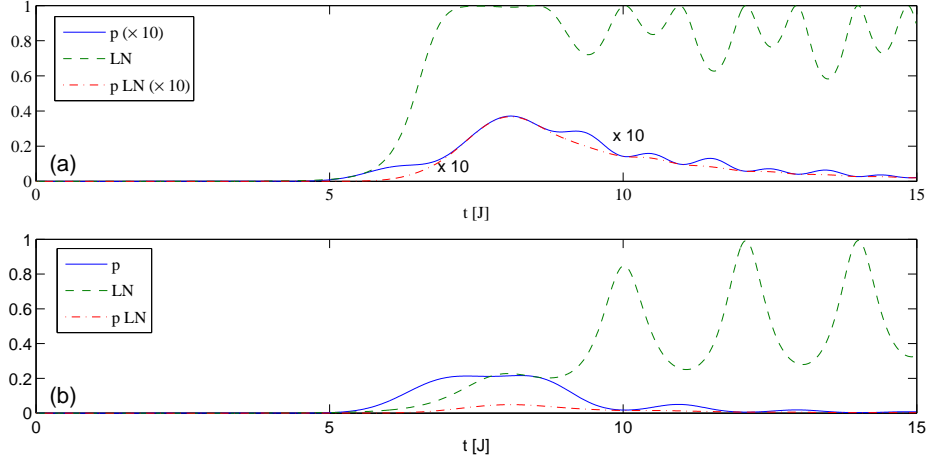}
\caption{Probability $p$ for the successful projection onto the subspace of one particle per spatially delocalized qubit (straight blue line) , logarithmic negativity ${\rm LN}$ (dashed green line) between the qubits after successful projection, and the product $p\;{\rm LN}$ (dashed red line) for a chain of $24$ sites with (a) no interaction $U/J=0$ and (b) strong interaction $U/J=20$. Results from MPS simulations with $D=18$.}
\label{fig:sqq}
\end{figure}
The probability $p$ of a successful projection onto the subspace of one particle per
SDQ is given by $p=\tr{\hat{\rho}^{\rm sdq}}\leq1$, and the entanglement between the two
spatially delocalized qubits in case of a successful projection is measured by the logarithmic negativity of
the correctly normalized state $\LN(\hat{\rho}^{\rm sdq}/p)$. The probability $p$, the logarithmic negativity $\LN$, and
the probabilistic entanglement $p\;\LN$ are plotted in Fig.~\ref{fig:sqq} (a) for the case of no interaction between the two bosons
($U/J=0$) and for strong interaction between them ($U/J\gg1$) in Fig.~\ref{fig:sqq} (b).

The time-dependence of the logarithmic negativity, as well as of the probability $p$, is clearly different in the two cases. It is illustrative to consider the case of only four sites, (with the two particles initially located at sites $2$ and $3$). Then the initial state is $\ket{10}_{\rm SDQ}$, which is a separable state of the two spatially delocalized qubits. If interactions are absent, then already at early times populations of states lying in the delocalized qubit space can originate from two possible paths (starting at sites $2$ or $3$, respectively). It is
this interference which leads to a fast generation of entanglement. If interactions are strong, then one of this paths is effectively suppressed
at early times as each particle is confined to ``its" qubit (this leads to the larger probability $p$ for a successful projection). Entanglement in this case is (initially) only generated through the collisional phase shift (this effect can be used to implement a phase gate for spatially delocalized qubits \cite{mompart:2003}), which however is small. As a consequence, entanglement is smaller if interactions are large.
\begin{figure}[t]
\centering
\includegraphics[width=0.75\linewidth]{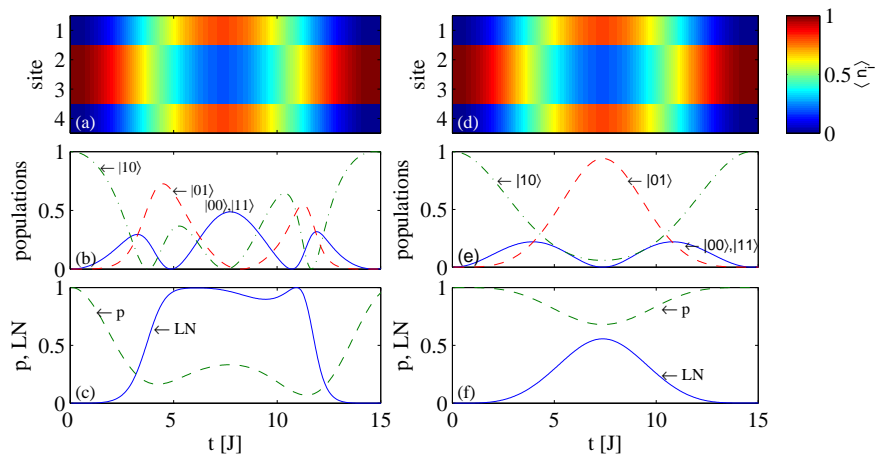}
\caption{
Generation of entanglement between two spatially delocalized qubits formed from sites $1,\,2$ and $3,\,4$, respectively.
Left column: $U/J=0$, right column: $U/J=20$. (a,d) show the mean site occupation
$\ew{n_i}$, (b,e) the populations ${}_{\rm SDQ}\bra{\alpha\beta}\hat\rho^{\rm SDQ}(t)/p\ket{\alpha\beta}_{\rm SDQ}$,
and (c,f) give the probability $p$ of successful projection into the subspace of one particle per spatially delocalized qubit
and the entanglement measured {\it via} the logarithmic negativity $\LN$.
}
\label{fig:sdq_4sites}
\end{figure}

In the presence of a ground state background with an average number $\bar n$ of particles per site,
generating (entangled) spatially delocalized quantum bits from the evolution of two extra particles only
can be done effectively in the Mott case. Here the definition of the basis can be modified as
$\ket{0}_{\rm SDQ}\equiv\ket{n_i=\bar n+1,n_{i+1}=\bar n},\,\ket{1}_{\rm SDQ}\equiv\ket{n_i=\bar n,n_{i+1}=\bar n+1}$.
In the superfluid case, the on-site particle number fluctuations in the ground state require a projection
onto the subspace of fixed number of particles {\it per} spatially delocalized qubit, which strongly reduces the
efficiency of the scheme.

%
%
%
\section{Conclusions}\label{sec:concl}

Summarizing, we have analyzed the generation of entanglement between the outer extremes of a 
1D Bose-Hubbard chain by loading an extra particle on top of the ground state. 
We have investigated effects arising from direct competition between tunneling and interactions in the entanglement behavior.
In some limiting cases, the bosonic propagation can be adequately described
as continuous time quantum walks. As part of our analysis, we have discussed  
two conceptually different ``computational bases'' to quantify entanglement between particles which have no internal 
degrees of freedom, namely the occupation number basis and a spatially delocalized qubit basis.

\section*{Acknowledgements}
 
We thank  M. Lewenstein, B. Paredes, and D. Porras for discussions. 
We acknowledge support from EU IP Programme ``SCALA'', 
European Science Foundation PESC QUDEDIS,
and MEC (Spanish Government) under contracts  AP2005-0595, FIS 2005-04627, FIS 2005-01369, EX2005-0830, CIRIT SGR-00185, CONSOLIDER-INGENIO2010  CSD2006-00019 ``QOIT".

\section*{References}

\end{document}